\documentclass{elsart}
\usepackage{graphicx}
\usepackage{amsmath}
\usepackage{amssymb}

\begin{document}

\begin{frontmatter}

\title{Spin-Spin Interaction In Artificial Molecules With In-Plane
Magnetic Field}

\author[add1,add2]{Devis Bellucci},
\author[add1,add2]{Massimo Rontani},
\author[add1,add2]{Guido Goldoni\thanksref{thank1}},
\author[add1,add2]{Filippo Troiani}, and
\author[add1,add2]{Elisa Molinari}
\address[add1]{INFM - National Research Center on nanoStructures
and bioSystems at Surfaces (S$^3$)}

\address[add2]{Dipartimento di Fisica, Universit\`a degli Studi di
Modena e Reggio Emilia, Via Campi 213/A, 41100 Modena, Italy}

\thanks[thank1]{Corresponding author. E-mail goldoni.guido@unimore.it}

\begin{abstract}
We investigate theoretically the spin-spin interaction of
two-electrons in vertically coupled QDs as a function of the angle
between magnetic field and growth axis. Our numerical approach is
based on a real-space description of single-particle states in
realistic samples and exact diagonalization of carrier-carrier
Coulomb interaction. In particular, the effect of the in-plane
field component on tunneling and, therefore, spin-spin interaction
will be discussed; the singlet-triplet phase diagram as a function
of the field strength and direction is drawn.
\end{abstract}

\end{frontmatter}

\section{Introduction}

Coupled quantum dots, also called Artificial Molecules (AM),
extend to the molecular realm the similarity between quantum dots
(QDs) and artificial atoms \cite{Rontani01,Pi01}. Inter-dot tunneling
introduces an energy scale which may be comparable to other energy
scales in the system, namely, single-particle confinement
energies, carrier-carrier interaction, and magnetic energy.

\begin{figure}
 \includegraphics[clip,width=35truemm,angle=-90]{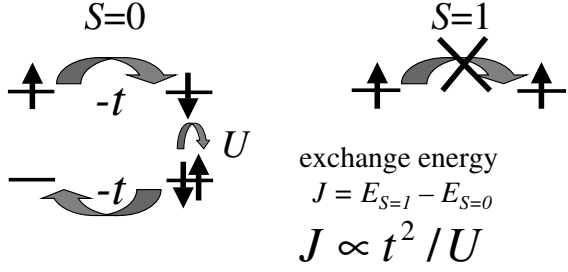}\\
  \caption{Sketch of the energy contributions to the
   tunneling-induced spin-spin interaction.}
  \label{fig:meccanismo}
\end{figure}

In AMs carriers sitting on either dot are not only
electrostatically coupled, but also have their spin interlaced
when tunneling is allowed \cite{Burkard00}. This is sketched in
Fig. \ref{fig:meccanismo}. For two electrons in a singlet state
it is possible to tunnel into the same dot. By doing so, they gain
the tunneling energy $t$; this may compensate for the loss in the
Coulomb energy $U$. This process is forbidden for two
electrons in the triplet state by Pauli blocking. Different spin orderings,
therefore, are associated to an exchange energy $J\propto t^2/U$.
While in real molecules $J$ is fixed by the bond length, in AMs it
is possible to tune all energy scales by sample engineering and
external fields.

One convenient way to control inter-dot tunneling, and, hence,
effective spin-spin interaction, is by applying a magnetic field
with a finite component perpendicular to the tunneling direction.
This is particularly important in vertically coupled QDs, where
otherwise tunneling in a given sample is fixed by sample
parameters; this extends the use of a vertical field, $B_\perp$,
to drive the system from a low correlation (low field) regime to a
high correlation (high field) one \cite{Rontani02}. In addition to
the vertical component of the field, therefore, a magnetic field
in the plane of the QD, $B_\parallel$, can be used to fully
control the spin-spin interaction and, therefore, the spin
character of the ground state of few-electron systems.

In this paper we discuss theoretically the two-electron phase
diagram, with particular respect to the spin ordering, in
vertically coupled QDs in the ($B_\perp$,$B_\parallel$) plane. Our
numerical approach is based on a real-space description of
single-particle states which fully includes the complexity of
typical samples, i.e., layer width, finite band-offsets etc. We
include carrier-carrier Coulomb interaction, represented in a
Slater determinant basis, by exact diagonalization methods.

\section{Single particle states}

\begin{figure*}
  \includegraphics[width=93truemm,angle=-90]{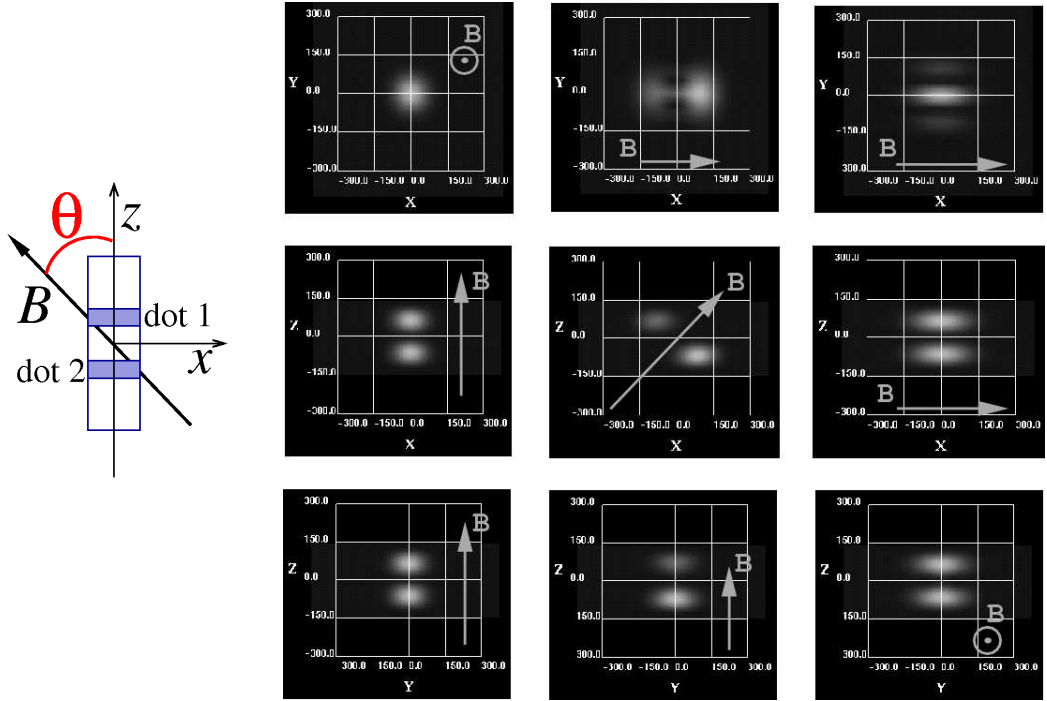}\\
  \caption{Single-particle ground state for a GaAs AM. The well width
of each dot along the growth direction is 10 nm, the inter-dot
potential barrier is 3 nm thick with an offset of 300 meV,
  and the lateral confinement energy $\hbar\omega_0 $ is 10 meV.
 Lengths are in \AA. From left to right column: magnetic field of 30 T at $0^\circ$,
 $45^\circ$, and $90^\circ$ with respect to the growth axis.
  The geometry and axis definition are sketched on the left.}
  \label{fig:single_particle}
\end{figure*}

In the symmetric coupled QD structure considered here (see sketch
in Fig. \ref{fig:single_particle}), zero field single-particle
orbitals consist of symmetric (S) and antisymmetric (AS) orbitals with
cylindrical symmetry due to a 2D harmonic confinement with a
characteristic energy $\hbar\omega_0$. A vertical field preserves
the symmetry, but splits the energy shells with non-zero angular
momentum ($p$, $d$, \ldots) and gives rise to the well known
Fock-Darwin level structure \cite{Jacak} which, in symmetric AMs, is
replicated for S/AS levels.

When a finite in-plane component of the field is applied by, e.g.,
rotating the sample with respect to the field, the energy spectrum
(not shown here) may be affected in two ways. First, anticrossing
between S and AS levels may occur depending on the angular
momentum component. Second, S/AS gaps may be reduced, indicating
that tunneling is suppressed by the in-plane field
\cite{Tokura00}.

 The effect of an in-plane field on the single-particle wavefunctions
 is exemplified in Fig. \ref{fig:single_particle} for a typical vertically
 coupled QD
structure. The field is increasingly rotated with respect to the
growth axis from vertical (left panels) to horizontal (right
panels) plane. In the top row the charge density is shown in the
plane perpendicular to the growth axis. When $B_\perp$ is switched
on by tilting the total field, the cylindrical symmetry of the
system is broken. Angular momentum is not a good
quantum number anymore, and energy shells with non-zero angular momentum
split and mix as the in-plane field increases. This results in
modulations in the charge density, as shown in Fig.
\ref{fig:single_particle}. In the middle and bottom row the charge
density is shown in two planes which contain the growth axis and
parallel (middle row) or orthogonal (bottom row) to the field. It
is shown that single particle states are modified in such a way
that the lobes of the wavefunctions located in either QD are
shifted in the plane identified by the field and the growth
direction, if both $B_\perp$ and $B_\parallel$ are non-zero. On
the other hand, if only one component of the field is present, the
$z\rightarrow -z$ symmetry is recovered, and the lobes are
vertically aligned. Moreover, comparing the left and right panels
one can see that the wavefunctions are squeezed in the direction
of the field, as expected. Therefore, a sufficiently large
in-plane field suppresses tunneling, and S and AS orbitals become
degenerate.

\section{Single-triplet transition}

We next consider the two-electron system. At low vertical fields
the ground state of single and coupled QDs is known to be a
singlet state \cite{Merkt91,Oh96}. In the moderate field regime,
therefore, the lowest energy levels are nearly unaffected by the
rotation except for the shift due to the reduction of the
tunneling energy, with the singlet state being the lowest.

\begin{figure}
  \includegraphics[width=76truemm]{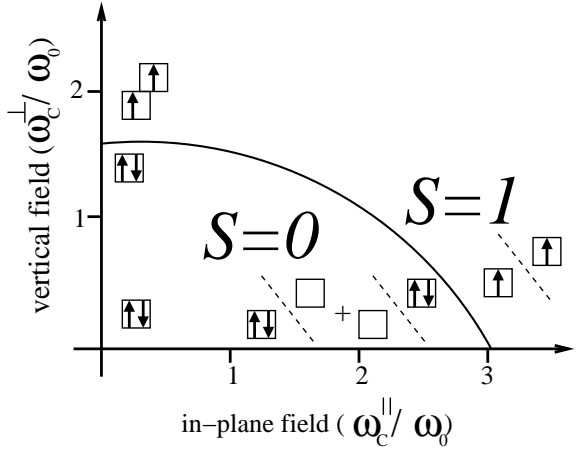}\\
  \caption{Single-triplet phase diagram calculated for a GaAs AM
  made of two QDs 10 nm wide along the growth direction, a 3 nm
  wide barrier, and lateral confinement energy
 $\hbar\omega_0 =4$ meV.
  The insets show the single-particle occupation in terms of
  S (left to the dashed lines) and AS orbitals (right to the dashed
  lines). At low $B_\parallel$ only S orbitals are occupied.
  $B$ is expressed in terms of the cyclotron frequency
  $\omega_c=eB/cm^*$, $m^*$ being the effective GaAs electron mass.
  }\label{fig:phase_diagram}
\end{figure}

At sufficiently high vertical field one or more (depending on the
sample parameter) singlet-triplet transitions take place at given
threshold fields, with the triplet state eventually being the
stable one. Since a finite $B_\parallel$ affects the tunneling
and, therefore, the exchange energy, the threshold fields will be
lowered as $B_\parallel$ increases. This is shown in Fig.
\ref{fig:phase_diagram}. The singlet state is stable in the low
field regime. The triplet state becomes favored in the large field
regime; is should be noted, however, that this happens by
different mechanisms whether $B_\perp$ or $B_\parallel$ is large.
In the former case, the squeezing of the wavefunction has a
Coulomb energy cost which can only be avoided by triplet spin
order. This is analogous to single QDs. However, while a finite
$B_\parallel$ would not affect very much electronic states in
single QDs, where single-particle gaps are large, in AMs the
in-plane field affects the S/AS gap; when this vanishes, no
tunneling energy is lost by paralleling the electron spins, and
the triplet state is favored.

In Fig. \ref{fig:phase_diagram} we also show in the insets the
character of the two-electron wavefunction of the ground state. In
the low $B_\parallel$ regime, the two electrons occupy only the S
state, either with the $s$ symmetry with opposite spin (low field)
or the $s$ and $p$ symmetry levels with the same spin orientation
(high field), since a large vertical field reduces the $s-p$ gaps.
In the large $B_\parallel$ regime, on the contrary, S and AS
states become degenerate, and are equally occupied by the two
electrons, due to Coulomb correlations.

\begin{figure}
  \includegraphics[width=76truemm]{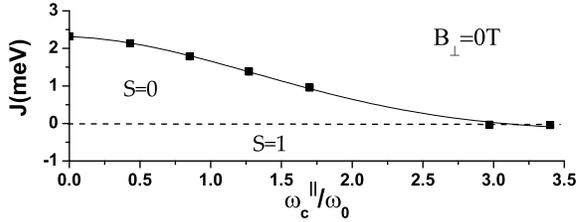}\\
  \caption{Exchange energy $J = E_{S=0}-E_{S=1}$ for a a GaAs AM.
 Same parameters as in Fig. \ref{fig:phase_diagram}.
}

  \label{fig:exchange_energy}
\end{figure}

Finally, in Fig. \ref{fig:exchange_energy} we show the exchange
energy at zero vertical field. This is positive (i.e., the singlet
is the ground state) at low fields, but rapidly decreases as the
field increases. At large fields, the exchange energy changes
sign, being eventually dominated by Zeeman energy.

\section*{Acknowledgements}

This work was supported by MIUR-FIRB ``Quantum phases of
ultra-low electron density semiconductor heterostructures''
and by INFM I.T. ``Calcolo Parallelo'' (2003).
Con il contributo del Ministero degli Affari Esteri,
Direzione Generale per la Promozione e la Cooperazione
Culturale.

\end{document}